\documentclass{webofc}
\usepackage[varg]{txfonts}   
\begin{document}
\title{Morphological analysis of SZ and X-ray maps of galaxy clusters with Zernike polynomials}
%
%

\author{\firstname{Valentina} \lastname{Capalbo}\inst{1}\fnsep\thanks{\email{valentina.capalbo@uniroma1.it}} \and 
	\firstname{Marco} \lastname{De Petris}\inst{1}
	\and
	\firstname{Federico} \lastname{De Luca}\inst{2}
	\and
	\firstname{Weiguang} \lastname{Cui}\inst{3}
	\and
	\firstname{Gustavo} \lastname{Yepes}\inst{4,5}
	\and
	\firstname{Alexander} \lastname{Knebe}\inst{4,5,6}
	\and
	\firstname{Elena} \lastname{Rasia}\inst{7,8}
	\and
	\firstname{Florian} \lastname{Ruppin}\inst{9}
	\and
	\firstname{Antonio} \lastname{Ferragamo}\inst{1}
}

\institute{Dipartimento di Fisica, Sapienza Università di Roma, Piazzale Aldo Moro 5, I-00185 Roma, Italy
\and
           Dipartimento di Fisica, Università di Roma 'Tor Vergata', Via della Ricerca Scientifica 1, I-00133 Roma, Italy
\and           
           Institute for Astronomy, University of Edinburgh, Royal Observatory, Edinburgh EH9 3HJ, UK
\and
           Departamento de Física Teórica, Módulo 8, Facultad de Ciencias, Universidad Autónoma de Madrid, E-28049 Madrid, Spain 
\and
           Centro de Investigación Avanzada en Física Fundamental (CIAFF), Facultad de Ciencias, Universidad Autónoma de Madrid, E-28049 Madrid, Spain
\and           
           International Centre for Radio Astronomy Research, University of Western Australia, 35 Stirling Highway, Crawley, Western Australia 6009, Australia
\and
           National Institute for Astrophysics, Astronomical Observatory of Trieste (INAF-OATs), Via Tiepolo 11, I-34131 Trieste, Italy
\and
           Institute for Fundamental Physics of the Universe (IFPU), Via Beirut 2, I-34014 Trieste, Italy
\and           
           Kavli Institute for Astrophysics and Space Research, Massachusetts Institute of Technology, Cambridge, MA 02139, USA
           }

\abstract{
Several methods are used to evaluate, from observational data, the dynamical state of galaxy clusters. Among them, the morphological analysis of cluster images is well suited for this purpose.
We report a new approach to the morphology, which consists in analytically modelling the images with a set of orthogonal functions, the Zernike polynomials (ZPs). We validated the method on mock high-resolution Compton parameter maps of synthetic galaxy clusters from {\sc The Three Hundred} project. To classify the maps for their morphology we defined a single parameter, $\mathcal{C}$, by combining the contribution of some ZPs in the modelling. We verify that $\mathcal{C}$ is linearly correlated with a combination of common morphological parameters and also with a proper 3D dynamical-state indicator available for the synthetic clusters we used. 
We also show the early results of the Zernike modelling applied on Compton parameter maps of local clusters ($z<0.1$) observed by the \textit{Planck} satellite. 
At last, we report the preliminary results of this kind of morphological analysis on mock X-ray maps of {\sc The Three Hundred} clusters.
}
\maketitle
\section{Introduction}
\label{intro}

Classify galaxy clusters based on their dynamical state is crucial to correctly infer other physical properties of those systems. For example, their mass can be estimated by exploiting the hypotheses of hydrodynamical and thermal equilibrium with an assumption of spherical distributions for both dark matter and baryonic components. However, in many cases they prove to be too simplified approximations. In fact, it is well known that galaxy clusters are dynamically active systems and their physical state does not always reflect a condition of equilibrium. Several studies are then focused on defining valuable methods to infer the real state of the clusters from observational data. Among the others, the analysis of the morphological appearance of multiwavelength images is a common approach used to classify clusters in different dynamical classes. Several morphological parameters can be defined based on characteristic features in the images and also used in some combinations (see e.g. \cite{Rasia2013}, \cite{Cialone2018}, \cite{DeLuca2021} and references therein).

Here, we report a new approach to the morphological analysis, developed in \cite{Capalbo2021}, which consists in modelling the cluster images with a set of functions, the Zernike polynomials (ZPs). This new method was validated, at first, on mock projections maps realized by exploiting the Sunyaev-Zel'dovich (SZ) effect \cite{SZ}, i.e. Compton parameter maps. We defined a single parameter to quantify the morphological differences between the cluster maps and we verified that it is well correlated with some common morphological parameters largely used in literature and also with a proper dynamical-state classification available from 3D data for the simulated clusters we used.
We also present the preliminary results of this method applied on the real Compton parameter maps of galaxy clusters from the \textit{Planck}-SZ catalogue and on mock X-ray maps.

\section{Analytical modelling with Zernike polynomials}
\label{ZPs}

The ZPs are a complete basis of orthogonal functions defined on a unit disk. Completeness and orthogonality make them well adapted to model functions or images in circular domains.
ZPs are defined as follows \cite{Noll1976}:
\begin{equation}
    \label{eq:Zpol}
    Z^m_n(\rho,\theta) = N^m_n R^m_n(\rho) \cos{(m\theta)}\,\,\, , \,\,\,
    Z^{-m}_n(\rho,\theta) = N^m_n R^m_n(\rho) \sin{(m\theta)}
\end{equation}
where $\rho$ is the normalized radial distance ($0\leq\rho\leq1$), $\theta$ is the azimuthal angle ($0\leq\theta\leq2\pi$), $n$ and $m$ are, respectively, the polynomial order and the angular frequency such that $m\leq n$ and $n-m$\,=\,even, $N^m_n=\sqrt{2(n+1)/(1+\delta_{m0})}$ is a normalization factor in which $\delta_{m0}$ is the Kronecker delta and $R^m_n(\rho)=\sum_{s=0}^{(n-m)/2} \frac{(-1)^s (n-s)!}{s!\Bigl(\frac{n+m}{2}-s\Bigr)!\Bigl(\frac{n-m}{2}-s\Bigr)!} \rho^{n-2s}$ is the radial term.
We emphasize that the goal of our work was not to reconstruct the cluster maps in details, but rather to reveal the main features able to identify different dynamical classes. For example, very regular patterns, mostly circular, can be related to relaxed systems, while asymmetries and substructures are signals of a disturbed dynamical state. In \cite{Capalbo2021} we analyzed mock Compton parameter maps at high resolution (see Sect.~\ref{datasets}) and within a circular aperture of radius equal to $R_{500}$, therefore we used ZPs up to the eight order $n$ (i.e. 45 terms, see the ordering scheme in \cite{Noll1976}) in order to obtain a spatial resolution of $\sim0.5R_{500}$ in the modelling (see also Appendix A in \cite{Capalbo2021}). We divided the ZPs in two classes, based on their 2D projections: terms with $m=0$, which show a circular symmetry and terms with $m\neq0$, which fit azimuthal inhomogeneities. We are currently using the same number of ZPs to model mock X-ray maps as well.
Each map is then expressed as a combinations of ZPs:
\begin{equation}
    \label{eq:fit}
    S_{y,X}=\sum_{n=0}^8 \sum_{m=0}^n c_{nm}Z_{nm}
\end{equation}
where $S_{y,X}$ indicates the spatial distribution of the Compton parameter $y$ or the X-ray surface brightness and $c_{nm}$ are the expansion coefficients of the single polynomials.
In the preliminary analysis on the \textit{Planck} Compton parameter maps we decided to reduce the number of ZPs in the fit since the resolution in the maps is significantly lower. Internal regions ($<R_{500}$) are poorly resolved, therefore we limit our modelling to a spatial resolution of $\sim R_{500}$, by using ZPs up to the fifth order (i.e. 21 terms).

\section{Data sets}
\label{datasets}

In this Section we describe three different types of maps used in our analyses: the mock Compton parameter and X-ray maps for synthetic galaxy clusters in the {\sc The Three Hundred} project \cite{Cui2018}, a large catalogue with 324 massive clusters ($M_{200}>6\times10^{14}h^{-1}$M$_{\odot}$ at $z=0$) generated through hydrodynamical simulations; the real Compton parameter maps of galaxy clusters from the \textit{Planck} observations.

\begin{itemize}
    \item {\bf Mock $y$-maps} Mock maps of the thermal component of the SZ effect \cite{SZ} were realized for all {\sc The Three Hundred} clusters. The maps are 2D projections of the Compton parameter $y$ (i.e. $y$-maps) which characterizes the effect. They were mimicked with the PyMSZ code\footnote{https://github.com/weiguangcui/pymsz} \cite{Cui2018}, with a spatial resolution of 10\,kpc per pixel corresponding to angular resolutions of $5.25^{\prime\prime}$, $1.14^{\prime\prime}$ and $0.59^{\prime\prime}$ respectively at $z=0$, 0.45 and 1.03, the reference redshifts analysed in \cite{Capalbo2021}. We normalized each map to the maximum and performed the Zernike fitting within a circular aperture of radius $R_{500}$ centred on the $y$-centroid. Note that the centroid was computed by fixing at first the circular aperture on the centre of the maps, which corresponds to the highest density peak.

    \item {\bf Mock X-ray maps} Mock X-ray maps were realized for {\sc The Three Hundred} clusters by using the PyXSIM code\footnote{https://github.com/jzuhone/pyxsim} \cite{ZuHone2016}. The maps are in terms of photon number counts, with 10 ks exposure time and with a fixed spatial resolution of 10\,kpc per pixel (same angular resolution as for the $y$-maps). They are related to the spectral band 0.2-15\,keV, convolved with the response file of the Wide Field Imager instrument for the \textit{Athena} satellite \cite{Meidinger2018}. We analysed them within a circular aperture of radius $R_{500}$ centred on the X-ray centroid, computed as in the mock $y$-maps. However, in this case we normalized the maps to the mean inside a region of radius $0.05R_{500}$, to smooth the contribution from saturated pixels.
    
    \item {\bf \textit{Planck} SZ-selected clusters} In the preliminary analysis of ZPs applied to real $y$-maps we selected a sample of clusters at $z<0.1$ from the PSZ2 catalogue \cite{PSZ2}. We used the public available full-sky $y$-maps \cite{Planck_ymaps} realized with two component separation algorithms: MILCA \cite{Hurier2013} and NILC \cite{Remazeilles2011}. These are maps in HEALPIX format with pixel size of $1.7^{\prime}$ and with a final angular resolution of $10^{\prime}$. Therefore, we only selected clusters with $R_{500}\leq10^{\prime}$ (i.e. resolved clusters). We extracted gnomonic projections from the HEALPIX $y$-maps, centred on the cluster coordinates and with a side-length of $2R_{500}$ for each cluster. We also used the available point source mask \cite{Planck_ymaps} to select maps with a low residual contamination from point sources, based on the spatial resolution of the Zernike modelling described in Sect.~\ref{ZPs}. Our final sample was composed by 135 galaxy clusters.
\end{itemize}

To estimate the capability of the Zernike morphological analysis in recognizing different dynamical classes in the \textit{Planck} cluster sample, we also used \textit{Planck}-like maps realized for {\sc The Three Hundred} clusters described above. For those clusters, in fact, we can evaluate \textit{a priori} the dynamical state from 3D data.
The mock \textit{Planck} maps were redone to mimic real \textit{Planck} maps in these following steps: at first the maps were convolved with a $10^{\prime}$ beam and gridded in pixels of $1.7^{\prime}$; a full-sky noise map was realized by using the \textit{Planck} noise power spectrum and the public available full-sky maps of the standard deviation of the Compton parameter; a patch was then extracted in this noise map at a random position (considering only the positions of detected clusters) to generate the final noise added in the $y$-maps. Note that the mock maps do not include point source contamination. We analysed them within a circular aperture of radius $R_{500}$, as for the real \textit{Planck} maps. In particular, we used the mock maps realized at four snapshots of redshift ($z=0.02, 0.04, 0.07, 0.09$) to cover the redshift range of the \textit{Planck} clusters.

\section{Results}
\label{results}

At first, we report the application of the Zernike fitting on cluster $y$-maps. We summarize the results in \cite{Capalbo2021} for mock $y$-maps and then the preliminary analysis on real maps for clusters in the PSZ2 catalogue. At last, we briefly discuss the ongoing analysis on X-ray maps.

\subsection{Zernike analysis for $y$-maps}

\begin{figure*}
\begin{center}
\includegraphics[scale=0.39]{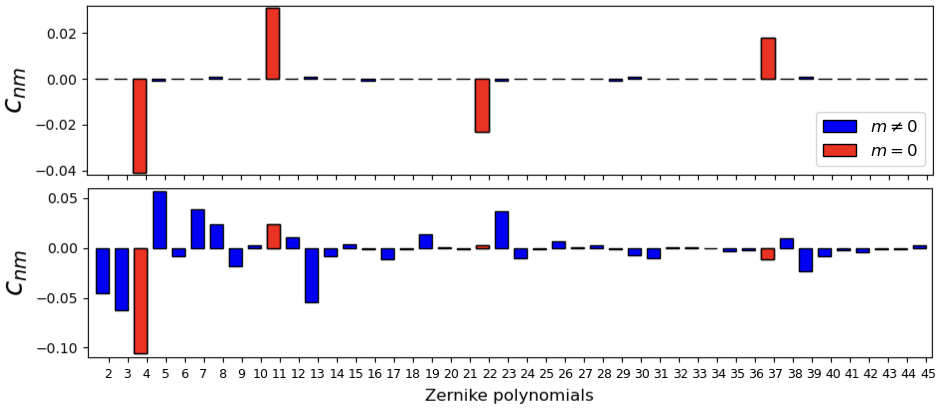}
\caption{Zernike coefficients $c_{nm}$ computed by fitting $y$-maps of a relaxed (top) and a disturbed (bottom) cluster. Red bars refer to ZPs with $m=0$, while blue bars are for $m\neq0$. The ZPs are ordered by following the Noll's scheme \cite{Noll1976}. Note that the first term ($Z_0^0$) is neglected because the corresponding $c_{00}$ is simply equal to the mean value of the signal inside the circular aperture in the map.}
\label{fig:Z_fit}       
\end{center}
\end{figure*}

The synthetic clusters we analysed in \cite{Capalbo2021} were previously classified in \cite{DeLuca2021} for their dynamical state by using a combination of some 3D dynamical indicators, i.e. the $\chi$ parameter (see also \cite{Haggar2020}). In addition, a morphological analysis of their $y$-maps were also performed in \cite{DeLuca2021} by applying a combined morphological parameter, $M$ (see also \cite{Cialone2018} for details on the parameters combined in $M$). Those results represented our reference to check the efficiency of the Zernike modelling.
When fitting the mock $y$-maps as in eq.~\ref{eq:fit}, we verified that ZPs with $m\neq0$ have $c_{nm}$ values negligible in case of regular (mostly circular) distributions in the maps, while they increase when dealing with complex patterns involving e.g. asymmetries or substructures. On the contrary, ZPs with $m=0$ have $c_{nm}$ values almost invariant when fitting different morphologies. This behaviour is clarified in Fig.~\ref{fig:Z_fit}, in which we show the results of the Zernike fitting for a relaxed and a disturbed cluster (as classified with $\chi$ and $M$).
Therefore, we defined a single parameter, $\mathcal{C}$, as follows:
\begin{equation}
    \label{eq:C}
    \mathcal{C}=\sum_{n,m\neq0} |c_{nm}|^{1/2}
\end{equation}
We verified that $\mathcal{C}$ is able to distinguish different morphologies, showing a linear correlation of $\sim78\%$ at $z=0$ with $M$, that is quite stable at higher $z$. The correlation with $\chi$ is lower, $\sim62\%$ at $z=0$, and it is mild descreasing along $z$. We concluded that the Zernike fitting is a valid method to evaluate the cluster morphology and infer the cluster dynamical state. In particular, the parameter we extracted shows similar performances with respect to a more complex combination of several parameters.

\begin{figure*}
\begin{center}
\includegraphics[scale=0.36]{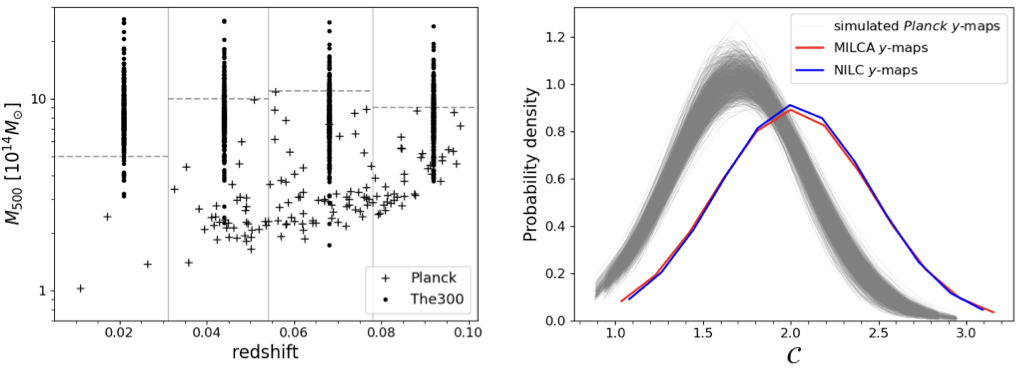}
\caption{Left: mass-redshift distribution of 135 \textit{Planck}-SZ clusters and of the synthetic clusters generated in The300 project at four redshifts. The vertical thin lines mark the redshift bins we used to compare the Zernike analysis on \textit{Planck} and mock (\textit{Planck}-like) $y$-maps. The horizontal dashed lines indicate the upper-mass limits defined to randomly extract the synthetic clusters in each bin.
Right: probability density for the distributions of $\mathcal{C}$ computed on mock $y$-maps of the synthetic clusters (grey) and on MILCA (red) and NILC (blue) $y$-maps of the \textit{Planck} clusters.}
\label{fig:Planck_M-z}       
\end{center}
\end{figure*}

For the current analysis of the \textit{Planck} $y$-maps we are also using \textit{Planck}-like maps generated for {\sc The Three Hundred} clusters in order to check the efficiency of the Zernike fitting in drawing a dynamical state evaluation for the \textit{Planck} cluster sample.
In Fig.~\ref{fig:Planck_M-z} (left panel) we show the cluster distribution in the mass-redshift plane, for both real and synthetic clusters.
In each redshift bin (marked by thin vertical lines in the figure) we perform random extractions of the same number of synthetic clusters as for the corresponding \textit{Planck} sub-sample, applying an upper-mass limit for the synthetic clusters defined by considering the value of the \textit{Planck} masses in each bin (see the horizontal dashed lines in the figure).
In this way we simulate several mock \textit{Planck} samples and for each one we compute the $\mathcal{C}$ parameter on all the $y$-maps. 
In the right panel in Fig.~\ref{fig:Planck_M-z} we show the distributions of $\mathcal{C}$ for the mock samples extracted (grey lines) and for the \textit{Planck} clusters (red and blue lines). The distributions in the two cases are quite different. However, we note that the \textit{Planck} clusters have, on average, lower masses with respect to the synthetic clusters (see left panel in Fig.~\ref{fig:Planck_M-z}). This is particularly evident at low redshift ($z\lesssim0.05$), where the number of \textit{Planck} clusters we selected is very low. This discrepancy could have an impact on the distributions we compared in the right panel in Fig.~\ref{fig:Planck_M-z}. Therefore, we plan to improve the analysis by considering a new sample of synthetic clusters which should be better representative of the \textit{Planck} mass distribution. 
However, in this first application we also performed a check to verify the correlation between $\mathcal{C}$ computed on the low-angular resolution ($10^{\prime}$) \textit{Planck}-like maps and the $\chi$ indicator defining the dynamical state of the synthetic clusters. From the random extractions described above we computed a mean (linear) correlation of $\sim46\%$, which we can consider as the efficiency of the Zernike fitting in evaluating the cluster dynamical state from \textit{Planck}-like maps.

\subsection{Zernike analysis for X-ray maps}

The ZPs method can also be applied to the X-ray maps. We used the mock X-ray maps from {\sc The Three Hundred} project to test that. In this case, the signal distribution in the maps varies on smaller spatial scales with respect to the $y$-maps (even if they have the same angular resolution, see Sect.~\ref{datasets}). This is due to the different dependence of the X-ray surface brightness ($\propto n_e^2$) and of the Compton parameter ($\propto n_e$) from the electron density $n_e$ in the intracluster medium. We found that the X-ray maps are poor modelled with the low-order ZPs we used for the $y$-maps. However, increasing the number of ZPs in the fit should be more expansive from a computational point of view, and without providing new key information about the cluster dynamical state. Therefore, we simply smoothed the maps by considering the logarithm of the photon number counts and we used 45 ZPs in the fit, the same as for the $y$-maps. Note that also these X-ray maps were analysed in \cite{DeLuca2021} by using the $M$ parameter. We obtained a good correlation at $z=0$ between $\mathcal{C}$ and both $M$ ($\sim80\%$) and $\chi$ ($\sim65\%$).
We also estimated a correlation of $\sim84\%$ between $\mathcal{C}$ computed on the X-ray maps and on the $y$-maps.
We plan to extend the X-ray analysis at higher redshifts and to compare in more details the results with the study of the $y$-maps in \cite{Capalbo2021}, in order to check if and how the morphological classification of the clusters changes in the two surveys.

\section{Conclusions}

The modelling of projection maps of galaxy clusters with ZPs results in a valuable approach to study their morphology and infer their dynamical state. By using a set of mock $y$ and X-ray maps we verified that it is possible to extract a single parameter from the Zernike fitting which is well correlated with combinations of more common morphological parameters or dynamical-state indicators. The advantage of this approach is its flexibility, i.e. the chance to simply tune the accuracy of the modelling by changing the number of ZPs used, by exploiting their completeness and orthogonality. This is particularly useful when dealing with real maps at low resolution and/or with a residual noise contamination. The method is well adapted to study clusters in large samples even if, in general, such a morphological analyses are limited by the angular resolution in the maps. We are currently using ZPs to study $y$-maps of local ($z<0.1$) \textit{Planck} SZ-selected clusters. The final results will be presented in an upcoming work.

%
%
%

\end{document}